\documentclass{article}
\usepackage{spconf,amsmath,graphicx}

\usepackage{algorithm}
\usepackage{algorithmic}
\usepackage{amssymb}
\usepackage{booktabs}
\usepackage{comment}
\usepackage{caption}
\usepackage{subcaption}
\usepackage{enumerate}
\usepackage{color}
\usepackage[dvipsnames, table]{xcolor}
\usepackage{colortbl}
\usepackage{multirow}
\usepackage{multicol}

\usepackage[pagebackref,breaklinks,colorlinks]{hyperref}

\definecolor{Gray}{gray}{0.95}
\definecolor{LightCyan}{rgb}{0.88,1,1}
\definecolor{Yellow}{rgb}{1,0.9,0.7}
\definecolor{Red}{rgb}{1,0.8,0.8}
\definecolor{Green}{rgb}{0.7,1,0.7}
\definecolor{Blue}{rgb}{0.8,1,1}

\newcommand{\RNum}[1]{\uppercase\expandafter{\romannumeral #1\relax}}

\newcommand{\tref}[1]{Table~\ref{#1}}

\newcommand{\ie}{\emph{i.e.} }
\newcommand{\eg}{\emph{e.g.} }
\newcommand{\etal}{\emph{et al.} }

\title{Toward Efficient Deep Blind RAW Image Restoration}
\name{Marcos V. Conde, Florin Vasluianu, Radu Timofte}
\address{Computer Vision Lab, CAIDAS \& IFI, University of Würzburg}

\begin{document}
%
\maketitle


\begin{abstract}
Multiple low-vision tasks such as denoising, deblurring and super-resolution depart from RGB images and further reduce the degradations, improving the quality. However, modeling the degradations in the sRGB domain is complicated because of the Image Signal Processor (ISP) transformations. Despite of this known issue, very few methods in the literature work directly with sensor RAW images. In this work we tackle image restoration directly in the RAW domain. We design a new realistic degradation pipeline for training deep blind RAW restoration models. Our pipeline considers realistic sensor noise, motion blur, camera shake, and other common degradations. The models trained with our pipeline and data from multiple sensors, can successfully reduce noise and blur, and recover details in RAW images captured from different cameras. To the best of our knowledge, this is the most exhaustive analysis on RAW image restoration. Code available at https://github.com/mv-lab/AISP
\end{abstract}


\begin{figure}[t]
     \centering

     \begin{subfigure}{\linewidth}
         \centering
         \includegraphics[width=\linewidth]{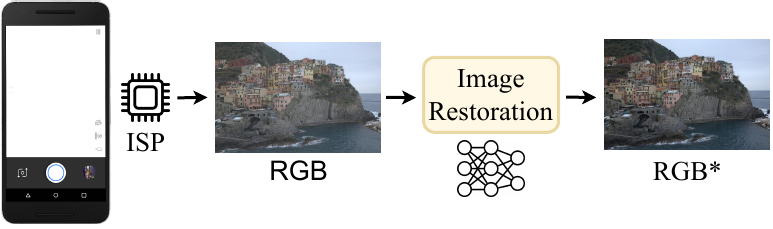}
         \caption{Image Restoration in the RGB domain as offline post-processing~\cite{zamir2022restormer, chen2022simple, conde2023perceptual} --- most popular setup and research.
         \vspace{2mm}}
         \label{fig:main-a}
     \end{subfigure}

     \begin{subfigure}{\linewidth}
         \centering
         \includegraphics[width=\linewidth]{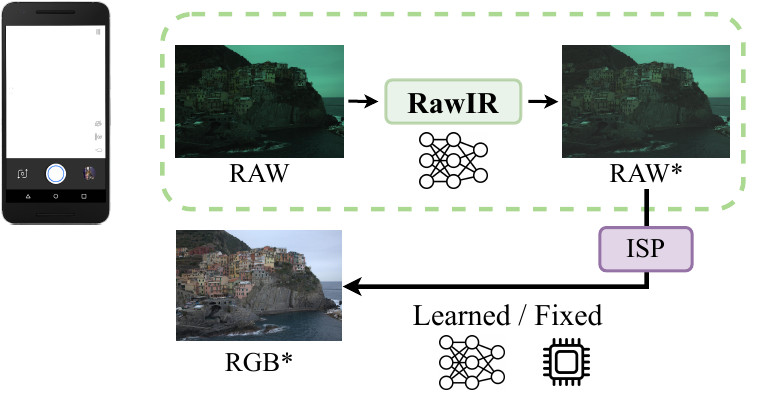}
         \caption{\textbf{Our approach RawIR}. We focus on RAW image restoration to benefit other downstream tasks. Moreover, RawIR can be plug \& play within the ISP, not only a post-processing as (a).}
         \label{fig:main-b}
     \end{subfigure}
     \caption{Comparison of image restoration pipelines. Note that previous approaches (a) depend on the in-camera ISP output, meanwhile, our RawIR method can complement any (learned) ISP, as the RAW restoration happens before it. We refer as RAW* and RGB* to the enhanced images. 
     }
     \label{fig:main-wild}
\end{figure}

\section{Introduction}
\label{sec:introduction}


The dedicated in-camera Image Signal Processor (ISP) converts RAW sensor data into less degraded and human-readable RGB images~\cite{schwartz2018deepisp,zhang2019zoom, conde2022modelbased}.

Photographers usually choose to process RAW images instead of RGB images to produce enhanced images with better perceptual quality, this is because of the RAW image advantages: 
(i) RAW data contains more information, typically 12-14 bits, whereas the ISP produced RGB is typically 8 bits. 
(ii) RAW data is linear \emph{w.r.t} the scene radiance, this facilitates modelling and correcting degradations such as noise and blur.
In contrast the ISP processing is highly non-linear and suffers an irreversible information loss throughout its multiple stages~\cite{karaimer2016software, brooks2019unprocessing}, which makes image restoration more difficult~\cite{delbracio2021mobile, xu2019rawsr}.
For these reasons, RAW images are a better choice than RGB images for many low-level vision tasks, such as image denoising and color constancy.


Despite these advantages, only a tiny portion of the low-vision literature works directly with RAW images. This is because the amount and availability of general RGB images (not sensor specific) is much larger than for their RAW counterparts. 
Therefore, the most relevant deep learning models for image restoration tasks use RGB images~\cite{zamir2022restormer, chen2022simple}. 

However such methods have important limitations. First and foremost, most methods are trained using synthetic data, and therefore they typically fail to generalize on real-world captures. Second, modelling complex degradations (and their interactions) for RGBs is difficult due to the non-linearities and information loss that happens at the ISP level.

The classical image degradation model is formulated as: 

\begin{equation}\label{eq:sisr_degradation}
  \mathbf{y}\!= \!(\mathbf{x}\otimes \mathbf{k})\! + \,\mathbf{n}
\end{equation}

It assumes the observed -or generated- degraded image $\mathbf{y}$ is obtained from an underlying clean image $\mathbf{x}$ by applying a degradation kernel (or PSF) $\mathbf{k}$, and the addition of the ubiquitous acquisition noise $\mathbf{n}$~\cite{hasinoff2014photon}. Therefore, the degradation pipeline is then characterized by the blur kernel $\mathbf{k}$ and noise level $\bf{n}$. The more realistic these factors are modelled and applied, the better the restoration models perform and generalize in real scenes~\cite{xu2019rawsr, kai2021bsrgan}.
Note that single image super-resolution (SISR) is considered a particular case of restoration, where besides the aforementioned degradations, we also assume some resolution downsampling~\cite{kai2021bsrgan}.
This introduces the need for more complex degradation pipelines, to reduce the generalization gap observed in the real world domain. Recently, \cite{kai2021bsrgan} designed a practical degradation pipeline for SISR in the RGB domain (Eq.~\ref{eq:sisr_degradation} adding downscaling), achieving promising results in real scene super-resolution \ie reducing noise, and improving sharpening and textures.

\vspace{2mm}

In this work we study \textbf{RAW image restoration} -see Fig. \ref{fig:main-b} and Eq.~\ref{eq:sisr_degradation}-, where all the information acquired by the camera is available, and the signal depends only on external factors (\eg light sources, user-camera interaction), and the quality of the used sensor. Moreover, the degradations do not suffer side-effects from the ISP stages~\cite{karaimer2016software, delbracio2021mobile}. Our realistic RAW degradation pipeline extends and improves previous methods~\cite{xu2019rawsr, xu2020exploiting} by integrating more diverse and realistic factors.

Our proposed model, RawIR, is an efficient blind RAW restoration model, trained using our degradation pipeline on real data from a family of smartphone Sony IMX sensors. The \textbf{main contributions} of this paper are:

\begin{itemize}
  \itemsep0em 
  \item[1)] We propose a pipeline to simulate degradations, which allows to train deep blind restoration models using RAW images. Our pipeline accounts for focusing issues, camera shake, motion blur, acquisition noise, and other related artifacts.
  \item[2)] A curated dataset for RAW image restoration, including a wide variety of camera sensors from smartphones. 
  \item[3)] A new benchmark for blind RAW image restoration -and reportedly the first one-. 
\end{itemize}

\section{Related Work}
\label{sec:related_work}

We discuss briefly \emph{state-of-the-art} image signal processing, degradation models, and deep blind restoration methods.

\subsection{Image Signal Processing}
\label{ssc:isp}

Smartphone and digital cameras have a dedicated in-camera Image Signal Processors (ISP) that process the RAW sensor readings captured by the camera into a less degraded and pleasant image to the human eye. 
Traditional ISPs usually apply a series of operations that include demosaicing $\uparrow_{\mathbf{dm}}$, white balance $\mathbf{W}$, color correction $\mathbf{C}$, tone mapping $\mathbf{\gamma}$, and others~\cite{karaimer2016software, delbracio2021mobile, conde2022modelbased}. The basic ISP function $f: \mathcal{X}_{raw} \rightarrow \mathcal{X}_{rgb}$ can be defined as:

\begin{equation}
\mathcal{X}_{rgb}\!= ~{\mathbf{\gamma}} \circ {\mathbf{C}} \circ {\mathbf{W}} \circ \uparrow_{\mathbf{dm}} \!(\mathcal{X}_{raw})
\label{eq:isp}
\end{equation}
where the RAW image follows a Bayer RGGB pattern as $\mathcal{X}_{raw} \in \mathbb{R}s^{\frac{H}{2} \times \frac{W}{2} \times 4}$, and $\mathcal{X}_{rgb} \in \mathbb{R}^{H \times W \times 3}$. The demosaicing step is highly related to denoising and super-resolution~\cite{qian2019trinity} --- thus, usually performed jointly.
Learning a faithful ISP transformation is a research topic by itself~\cite{schwartz2018deepisp,zhang2019zoom}, yet it is not the main focus of this work. 

\vspace{-2mm}
\subsection{Degradation Models}
\label{ssc:degradation_models}

As previously mentioned, designing realistic degradation models for data synthesis is a core task in low-level vision. The most popular degradation models usually consist of a sequence of blur and noise (and downsampling, in the case of SISR). Since most degradation models are far from the real-world complexity, some works aimed at solving such problem by designing complex degradation pipelines~\cite{xu2019rawsr}.

RAWSR~\cite{xu2019rawsr, xu2020exploiting} apply a single image super-resolution degradation pipeline for synthesizing realistic low-resolution RAW images and train models. They consider a heteroscedastic Gaussian noise and modest motion blur.

There is little study about more complex degradations beyond noise~\cite{hasinoff2014photon, liang2020raw} in the RAW domain. We improve these approaches by considering more realistic noise profiles, anisotropic defocus and motion blur, and exposure issues. Our degradation model is explained in Section~\ref{sec:method}.

\vspace{-2mm}
\subsection{Deep Blind Image Restoration}
\label{ssc:flexible_SISR}

In the recent years blind image restoration using deep learning has become a popular task~\cite{kai2017dncnn, nah2017Gopro, zhang2021DPIR, kai2021bsrgan, zamir2022restormer, chen2022simple}, this is because blind methods do not require prior knowledge about the degradation factors or the camera sensor, therefore they are general purpose models. For these reasons we also focus on the deep blind restoration problem, yet, in the RAW domain.
The vast majority of deep blind image restoration and SISR methods operate in the RGB domain. 

\vspace{2mm}

In comparison, there are far fewer works that solve \textbf{deep blind restoration for RAW images}.
Most works focus on RAW denoising, since estimating and correcting the noise on unprocessed raw data is well-studied~\cite{chen2018learning, abdelhamed2018high, brooks2019unprocessing, zamir2020cycleisp}. 
Liang~\etal studied RAW image deblurring on a simple setup using images from one DSLR camera~\cite{liang2020raw}. 
RawSR~\cite{xu2019rawsr, xu2020exploiting} for super-resolution and enhancement using both RAW and RGB images as inputs, and BSRAW~\cite{conde2024bsraw} using only RAW images.
%
\vspace{1mm}


\begin{figure*}[!ht]
    \centering
    \setlength\tabcolsep{1.0pt}
    \begin{tabular}{ccccc}
    \includegraphics[width=0.19\linewidth]{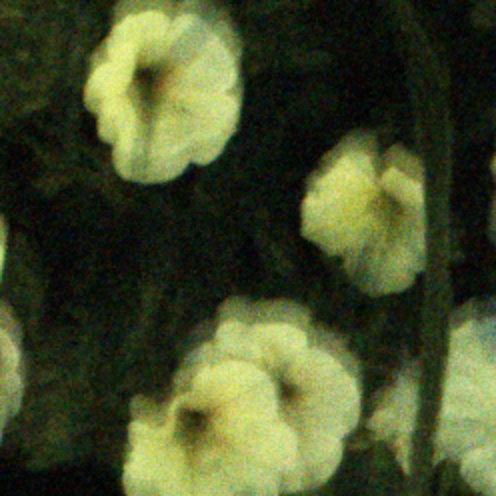} &
    \includegraphics[width=0.19\linewidth]{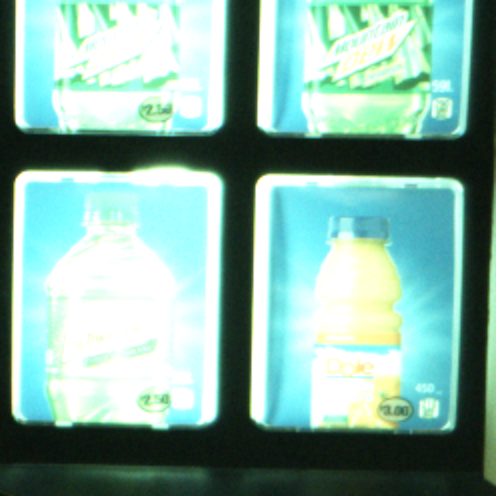} & 
    \includegraphics[width=0.19\linewidth]{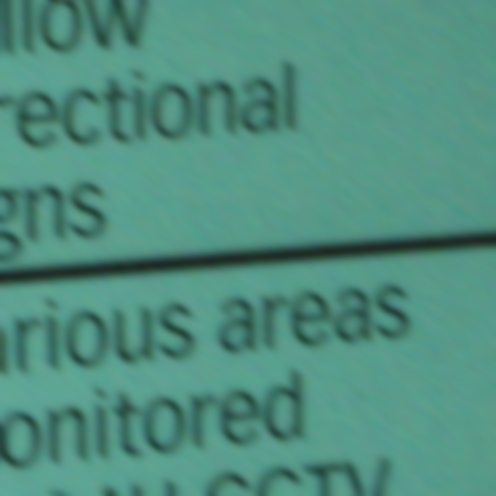} &
    \includegraphics[width=0.19\linewidth]{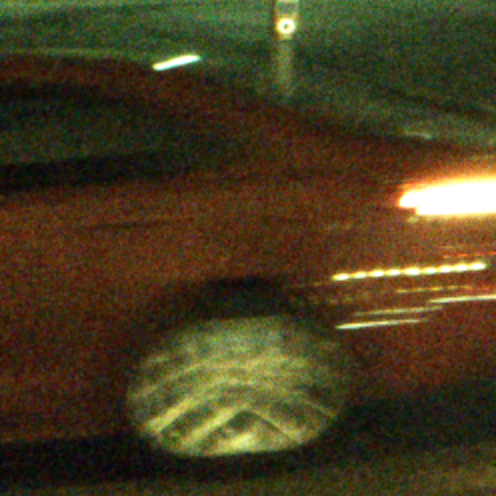} &
    \includegraphics[width=0.19\linewidth]{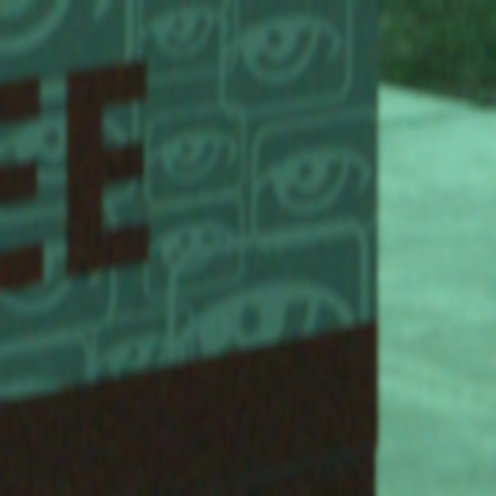}
    \tabularnewline
    \includegraphics[width=0.19\linewidth]{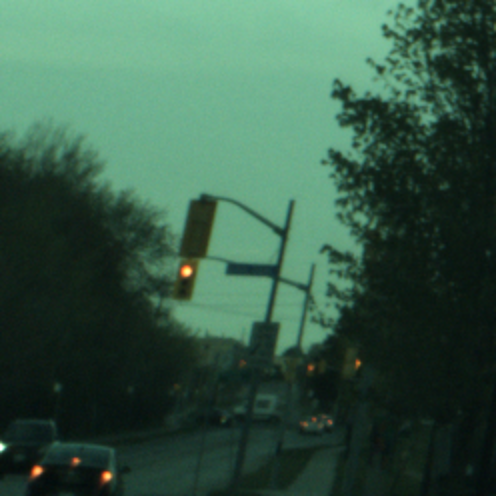} &
    \includegraphics[width=0.19\linewidth]{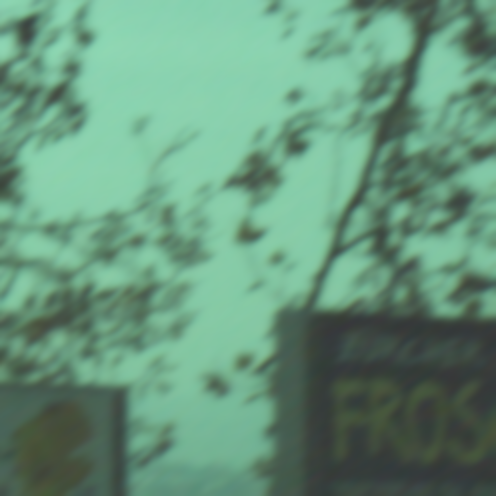} & 
    \includegraphics[width=0.19\linewidth]{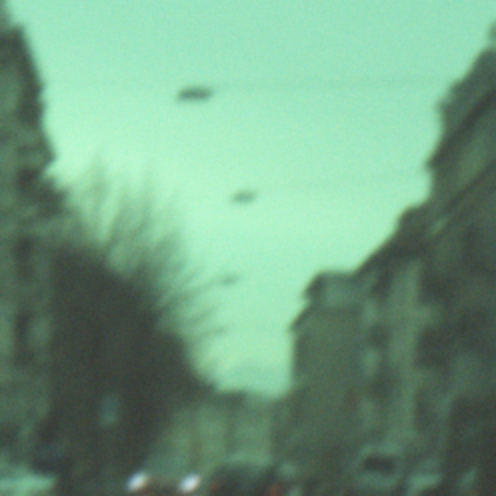} &
    \includegraphics[width=0.19\linewidth]{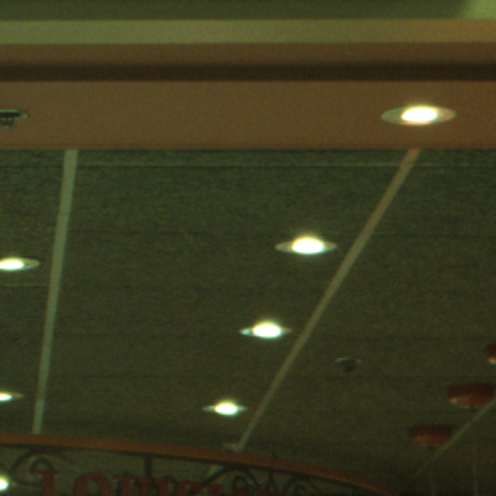} &
    \includegraphics[width=0.19\linewidth]{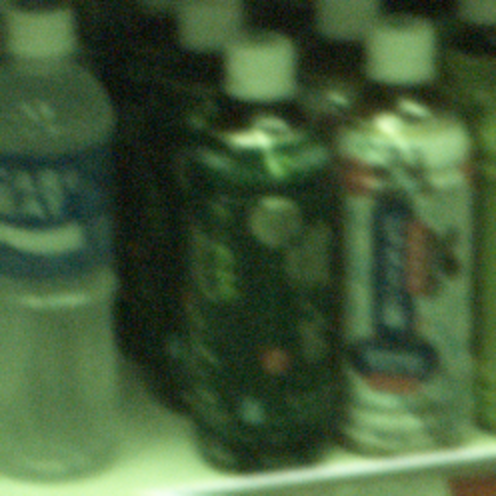}
    \tabularnewline
    \end{tabular}
    \caption{Degraded samples from our datasets. Five real, and five generated. To avoid biases, we do not indicate if the images are real or synthetic. We can appreciate defocus, motion blur, and noise among the most common degradations. The synthetic training dataset covers these degradations thanks to our degradation pipeline. Best viewed in electronic version.
    }
    \label{fig:deg_samples}
\end{figure*}


To the best of our knowledge, there is no benchmark for general RAW image restoration. In this work we build a \emph{new benchmark} considering state-of-the-art methods, and a curated dataset using multiple smartphone camera sensors and our new realistic degradation model. 
We aim to study this restoration problem assuming (i) unknown real-world degradations, and (ii) an unknown device \ie our method can be applied to images from different sensors, as the most popular RGB restoration models~\cite{kai2021bsrgan, zamir2022restormer,chen2022simple}. This sets a novel study in the field.

\section{Simulating RAW Degradations}
\label{sec:method}

We propose a new degradation model to synthesize realistic degraded RAW images for training deep blind restoration models. We address a wide variety of limiting factors related to the acquisition and the processing of high quality and high resolution RAW images: (i) noise -- related to the size of the used sensor and exposure time~\cite{karaimer2016software}. (ii) exposure. (iii) motion and defocus blur -- related to stabilization issues. (iv) LDR due to limited signal representation. 
Note that despite the promising results, previous methods~\cite{xu2019rawsr, xu2020exploiting} do not consider exposure, multiple noise profiles, or complex blur and PSFs. We provide multiple examples generated using our degradation pipeline in Figure~\ref{fig:deg_samples}.

\subsection{Noise}
\label{section:noise}

Noise ($\mathbf{n}$) is present in any image acquisition and processing pipeline. This depends on multiple factors \eg sensor size, exposure time, light conditions. In photography, the noise is typically removed -denoised- in the RAW domain, where it is well-studied due to its linear properties~\cite{hasinoff2014photon, hasinoff2016burst, abdelhamed2018high, brooks2019unprocessing} and easier to treat before applying the ISP non-linear steps~\cite{karaimer2016software}.
Most methods use Homoscedastic Gaussian noise~\cite{kai2021bsrgan, kai2017dncnn, zhang2021DPIR} in their degradation models. Other use sharper Heteroscedastic Gaussian distribution~\cite{abdelhamed2018high, xu2019rawsr}.

We adopt a more practical shot-read noise~\cite{hasinoff2014photon, brooks2019unprocessing, zhang2021rethinking}. In Eq.~\ref{eq:nat_noise}, we can observe the intensity $y$ as a sample of a Gaussian distribution having as mean the input signal $x$ and the variance depending on the $\lambda_{s}$ and $\lambda_{r}$ parameters~\cite{brooks2019unprocessing}. This is derived from a Poisson-Gaussian noise model:

\begin{equation}
y \sim \mathcal{N}(\mu=x, \sigma^2=\lambda_{r} + \lambda_{s}x)
\label{eq:nat_noise}
\end{equation}

We estimate noise profiles from our devices using Zhang~\etal method~\cite{zhang2021rethinking}, and we also use the popular noise profiles from SIDD~\cite{abdelhamed2018high, Abdelhamed_2019_CVPR_Workshops}for smartphones. Therefore, in comparison to previous approaches that use a single distribution~\cite{xu2019rawsr, brooks2019unprocessing}, our pipeline integrates \emph{multiple real noise profiles} that happen randomly at a time. We also simulate the interactions (\eg reduction, magnification) between the acquired noise and other degradations (\eg blur) by randomly choosing the point where the noise is injected.

\subsection{Blur}\label{section:blur}

Blur ($\mathcal{B}_\mathbf{k}$) is together with noise one of the most common degradations appearing in image acquisition \eg, motion blur, handheld camera shake, and defocus blur~\cite{hosseini2019convolutionalblur}. Capturing aligned blurry-clean pairs in real scenarios is extremely difficult, for this reason, the most popular deblurring datasets are synthetic (\eg GOPRO dataset of Nah~\etal~\cite{nah2017Gopro}).
Most approaches adopt a uniform blur by convolving the image with iso/anisotropic Gaussian kernels~\cite{kai2021bsrgan}. Xu~\etal~\cite{xu2019rawsr, xu2020exploiting} implemented defocus blur as a disk kernel, and a modest motion blur.

We create our blur degradation by convolving the image with a \emph{diverse pool of kernels}: classical isotropic and anisotropic gaussian blur kernels, real estimated motion blur kernels, and real estimated PSFs (point-spread-function) from~\cite{kai2021bsrgan}.
Each kernel in the pool is applied randomly. Considering the $xOy$ domain as the surface of the simulated sensor, we couple the $(\Sigma_x, \Sigma_y)$ variance parameters pair with the degree of instability suffered by the sensor during image acquisition along each axis. We uniformly sample the variance interval for both directions, therefore the reach of the kernel also varies. We apply maximum two kernels at a time. Since the order of these is stochastic, we can address a large variety of situations and the interactions with the noise model through the instability of the sensor. Note that in our test dataset, we do cover some of the most complex and common blur degradations.

\subsection{Exposure}

Exposure ($\mathcal{E}$) may become an unpleasant effect in photography. Either underexposure or overexposure can lead to noise or losing signals in the image~\cite{chen2018learning, hasinoff2016burst}. By integrating this degradation we do not aim to solve HDR~\cite{hasinoff2016burst}, but enabling the model to learn certain exposure corrections. 
We model low exposure with intensity rescaling using a linear ISO model to produce the resulting dimmed image. 

\subsection{Reduced Representation}

The number of bits used for the signal representation greatly impacts the quality of the image, with higher number of bits enabling more colors, HDR, and details in the acquired images. 
We follow a simple LDR-HDR synthesis method from~\cite{liu2020single}, and include a random uniform bitrate re-quantization. This enforces the model to generalize across sensors (\eg smartphones with 10-12 bits signal).
This operation is represented as $\Phi$, and affects maximum to 2 bits.

\begin{figure*}[!ht]
    \centering
    \setlength\tabcolsep{1.0pt}
    \begin{tabular}{c cccc}
    \raisebox{2cm}{\multirow{2}{*}{\includegraphics[width=0.21\linewidth]{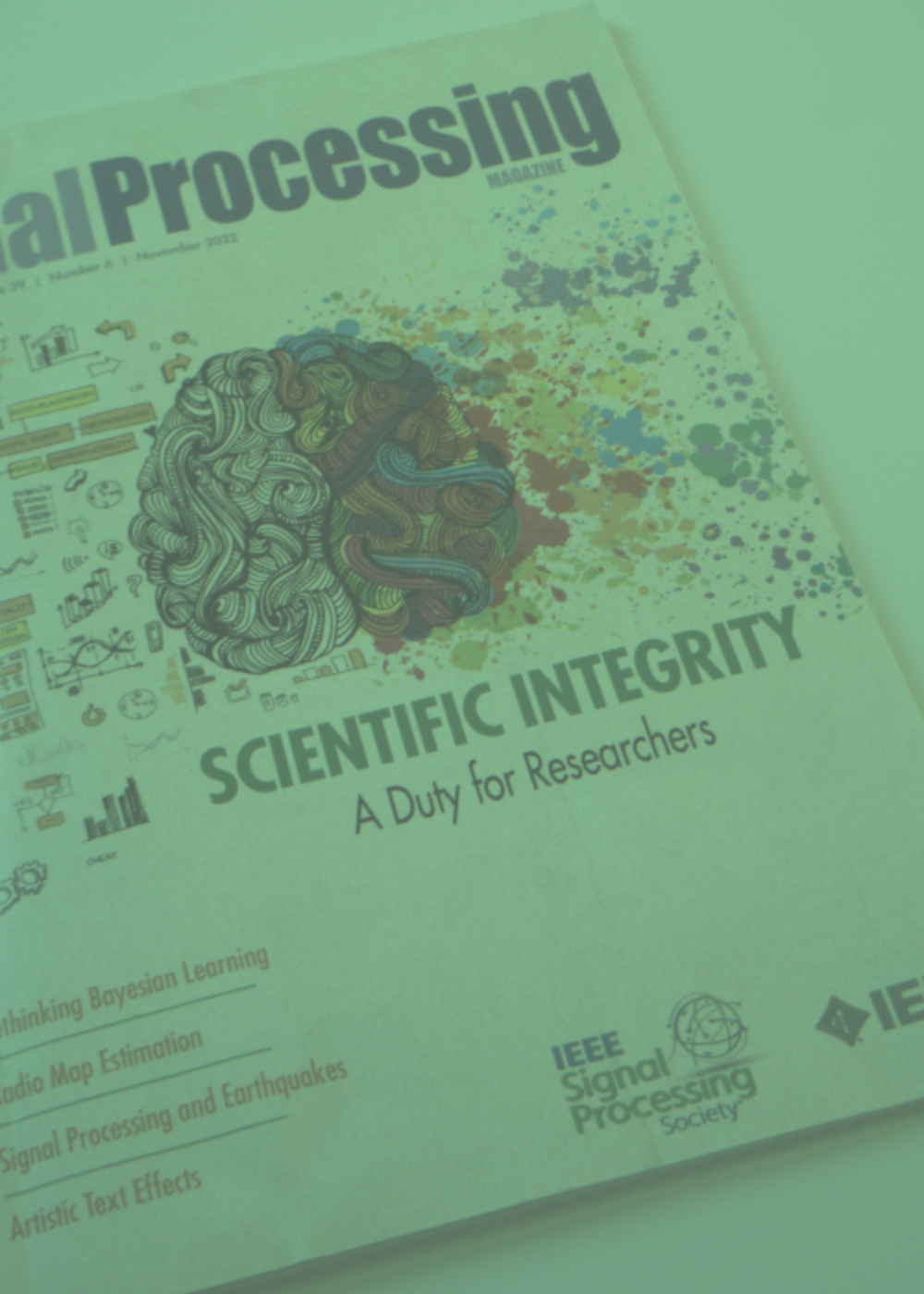}}} 
    \hfill
    &
        \includegraphics[angle=90,width=0.19\linewidth]{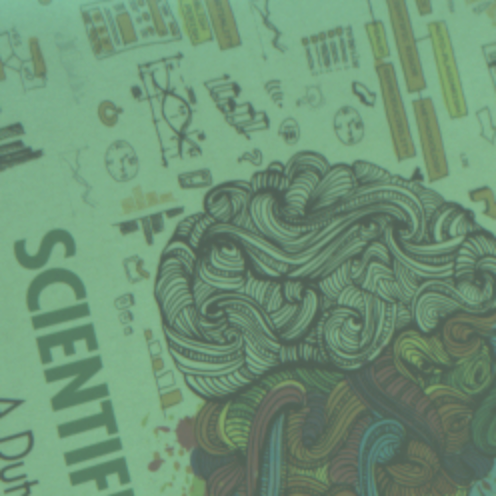}& 
        \includegraphics[angle=90,width=0.19\linewidth]{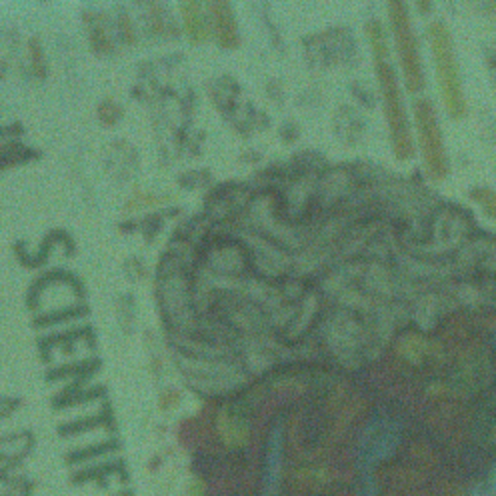} &
        \includegraphics[angle=90,width=0.19\linewidth]{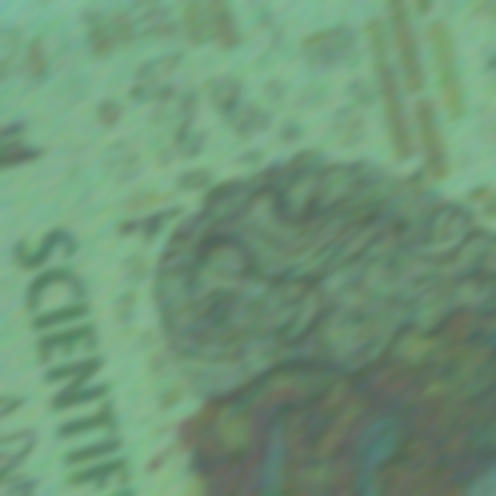} &
        \includegraphics[angle=90,width=0.19\linewidth]{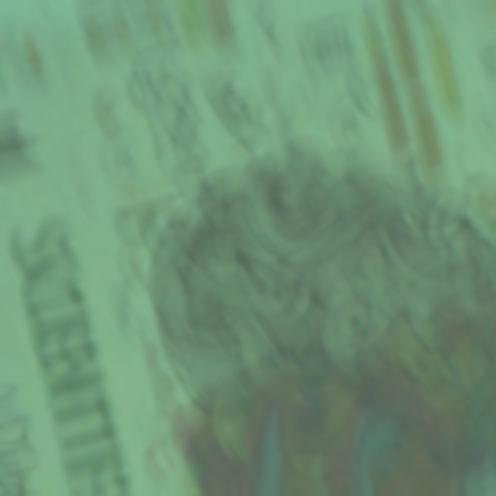}
        \tabularnewline
        & Input Raw $\mathbf{x}$ & 
        $\!\mathcal{E}((\mathbf{x} + \,\mathbf{n}) \otimes \,\mathbf{k})$ & 
        $\!\Phi(\mathcal{E}((\mathbf{x}+ \,\mathbf{n})\otimes \mathbf{k}))$ & 
        $\!(\mathbf{x} + \,\mathbf{n})\otimes \mathbf{k_1}\otimes \mathbf{k_2}$
        \tabularnewline
        %
        & \includegraphics[angle=90,width=0.19\linewidth]{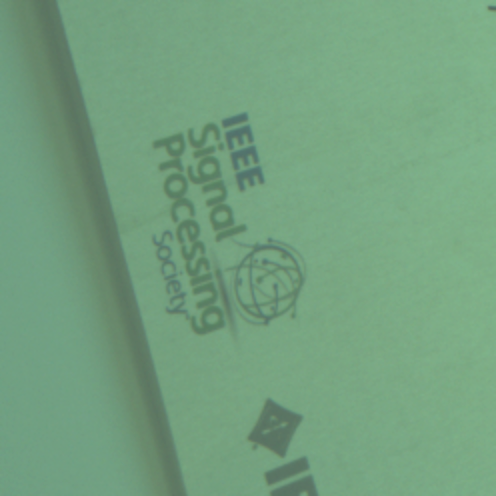} & 
        \includegraphics[angle=90,width=0.19\linewidth]{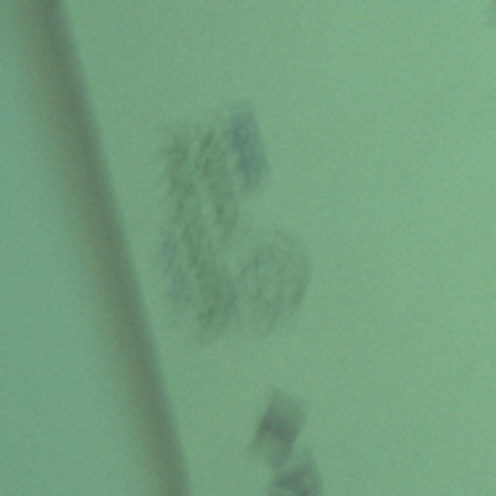} 
        &
        \includegraphics[angle=90,width=0.19\linewidth]{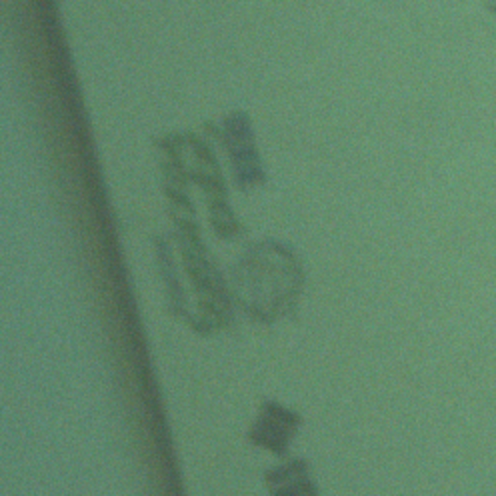} &
        \includegraphics[angle=90,width=0.19\linewidth]{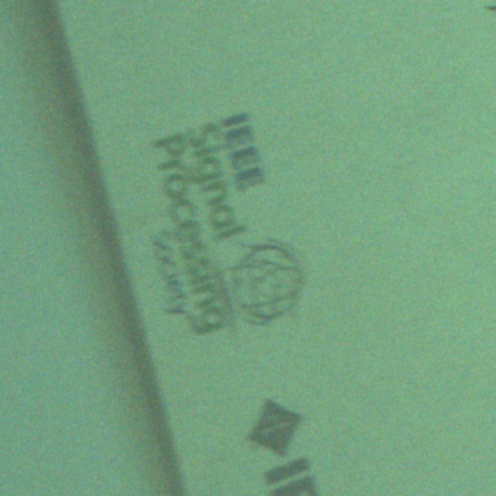}
        \tabularnewline
        HR Raw & Input Raw $\mathbf{x}$ & 
        $\!(\mathbf{x} \otimes \mathbf{k_1} \otimes \mathbf{k_2}) + \,\mathbf{n}$ & 
        $\!\Phi((\mathcal{E}(\mathbf{x}) + \,\mathbf{n})\otimes \mathbf{k})$ & 
        $\!(\mathbf{x} \otimes \mathbf{k}) + \,\mathbf{n}$
        \tabularnewline
    \end{tabular}
    \caption{Samples generated using our RAW degradation pipeline. For a reference input RAW image, a degradation sequence is performed to produce the corresponding degraded image (see Sec.~\ref{ssc:summary}). Thus, we can generate paired degraded/clean images for training deep blind restoration model. Each degradation is applied on the 4-channel RAW image~\cite{liu2019learningrawaug} respecting the RGGB pattern. RAW images are visualized through bilinear demosaicing. Image from our dataset.}
    \label{fig:degradation-pipe}
\end{figure*}

\subsection{Summary}
\label{ssc:summary}

Given the current description of the degradation pipeline, we covered most of the factors affecting the quality of RAW images. Considering that the injected noise $\mathbf{n}$ is sampled ($\leftarrow$) from multiple real noise profiles, and the kernels $\mathbf{k_i}, \mathbf{k_j}$ are sampled from a diverse pool of PSFs, we can update Eq.~\ref{eq:sisr_degradation} into Eq.~\ref{eq:final-model}. 
The typical and initial order of the operations is to apply the PSF before adding noise to the raw signal following the optics and image formation theory, as show in Eq.~\ref{eq:final-model}.

\begin{align}\label{eq:final-model}
\mathbf{y}\! = \!\Phi(\mathcal{E}((\mathbf{x} \otimes \mathcal{B}_\mathbf{k_i} \otimes \mathcal{B}_\mathbf{k_j}) + \,\mathbf{n} )) \\
\mathbf{n} \leftarrow \{\mathbf{n_1}, \mathbf{n_2} \dots \mathbf{n_m}\}\nonumber\\
\mathbf{k_i}, \mathbf{k_j} \leftarrow \{\mathbf{k_1}, \mathbf{k_2} \dots \mathbf{k_p}\} \nonumber
\end{align}

In Table~\ref{tab:deg} we summarize the two degradation levels used in our experiments: \RNum{1}) classical image restoration. \RNum{2}) Our proposed degradation model with more complex interactions.

\begin{table}[]
    \centering
    \begin{tabular}{c l l}
         \toprule
         Level & Degradations & Model \\
         \midrule
         \RNum{1} & $\mathbf{n}$, $\mathcal{B}_\mathbf{k}$ & $\mathbf{y} = \!(\mathbf{x}\otimes \mathbf{k}) + \,\mathbf{n}$ \\
         \RNum{2} & $\mathbf{n}$, $\mathcal{B}_\mathbf{k}$, $\mathcal{E}$, $\Phi$ & $\mathbf{y} = \!\Phi(\mathcal{E}(\mathbf{x}\otimes \mathbf{k} + \,\mathbf{n}$))  \\
         \bottomrule
    \end{tabular}
    \caption{Summary of the different levels of degradations included in our pipeline. Notation from Section~\ref{sec:method}.
    }
    \label{tab:deg}
\end{table}

\section{Discussion}
\label{sec:discussion}

Our proposal provides several advantages over previous methods~\cite{xu2019rawsr, xu2020exploiting}, allowing to synthesize realistic degraded RAW images for training models. We consider necessary to add discussion to further understand the potential of our contribution: \textbf{(i)} most related works evaluate their models on synthetic degraded data~\cite{nah2017Gopro, xu2019rawsr, qian2019trinity, kai2021bsrgan}. \textbf{(ii)} This degradation strategy is very practical, we can produce unlimited perfectly aligned training images --- in many cases obtaining aligned degraded-clean pairs is practically impossible~\cite{nah2017Gopro}. \textbf{(iii)} Training on synthetic data and generalize to real world scenes is possible if the adopted degradation pipeline is realistic~\cite{kai2021bsrgan, nah2017Gopro}. \textbf{(iv)} We do not aim at solving a particular degradation for a specific dataset \ie denoising at SIDD~\cite{abdelhamed2018high} or DND~\cite{plotz2017benchmarking}, as this is already well-studied.

We aim to develop \textit{general-purpose} blind restoration models for RAW images --- our restoration model can be applied to different related sensors \eg Sony IMX family. 

\textbf{(v)} the degradation pipeline can produce some extreme cases that rarely happen in real-world scenarios, while this can improve the generalization ability of the models~\cite{kai2021bsrgan}.
\textbf{(vi)} a single deep neural network with large capacity has the ability to handle different degradations~\cite{kai2021bsrgan, xu2019rawsr}.

Finally, the pipeline is designed to be fast and memory efficient, so it is integrated in our dataloader. This limits the scope of the applied operations (\eg we cannot use segmentation/semantic masks or multiple shots). 

\section{Experiments}

\subsection{RawIR Dataset}
\label{sec:datasets}

Our sources of real RAW data and our train/test split are summarized in Table~\ref{tab:datasets}. We use diverse \emph{smartphone data}. Since the RAW images contain certain degradations due to their ``raw'' nature, for the particular case of smartphones this is enhanced due to the small sensor~\cite{delbracio2021mobile}. We use smartphone images due to their complexity, variety and applications \emph{w.r.t} DSLR datasets. The RAW images are captured using diverse sensors from the same ``family": Sony IMX CMOS (2018-2020) --- up to 50MP.

\begin{table}[t]
    \centering
    \resizebox{\columnwidth}{!}{
    \begin{tabular}{l c c c c c}
         \toprule
         Device & Type & Sensor & \# Train  & \# Test & Res.\\
         \midrule
         Huawei P20 Pro   & B & IMX600   & 6000  & 500  & 248p \\
         iPhone X~\cite{afifi2021raw2raw}         & B & IMX333   & 8000  & 800  & 248p \\
         Samsung S9~\cite{afifi2021raw2raw}       & B & IMX345   & 8000  & 800  & 248p \\
         \midrule
         \rowcolor{Gray} Sony Xperia Z    & Q & IMX586    & 100     & -  & 448p \\
         \rowcolor{Gray} OnePlus Nord 2   & Q & IMX766    & 280     & -  & 448p \\
         \bottomrule
    \end{tabular}}
    \caption{Dataset split. The sensors highlighted in gray correspond to our collected images. We indicate the type of sensor, (B) Bayer and (Q) Quad, for the last one, we obtained the RAW images after remosaicing. We indicate the resolution (Res.) of the patches extracted from the full images. We use public ZRR~\cite{ignatov2020replacing} and Raw2Raw~\cite{afifi2021raw2raw}.
    }
    \label{tab:datasets}
\end{table}

The data \emph{pre-processing} is as follows: (i) we normalize all RAW images depending on their black level and bit-depth. (ii) we convert (``pack") the images into the well-known RGGB Bayer pattern (4-channels), which allows to apply the transformations and degradations without damaging the original color pattern information. (iii) Following previous work on RAW processing and learned ISPs~\cite{xu2019rawsr}, we train using image \emph{patches} of dimension $248\times248\times4$. 

We provide qualitative samples of our dataset (synthetic and real scenes) in Figure~\ref{fig:deg_samples}. Our RAW degradation pipeline can emulate realistically the most common degradations.

\vspace{2mm}
\noindent\textbf{Training and Testing.~} 
We apply our complete degradation pipeline (see Sec.~\ref{ssc:summary}) to the training images to generate aligned degraded-clean pairs. The \emph{test dataset} is generated by applying our degradation pipeline, at different levels, to all the corresponding clean test images -- see Table~\ref{tab:datasets}.

\begin{table}[t]
    \centering
    \resizebox{\linewidth}{!}{
    \begin{tabular}{l c c c c c c}
        \toprule
        & \multicolumn{2}{c}{Efficiency Metrics} & \multicolumn{2}{c}{Level~\RNum{1}} & \multicolumn{2}{c}{Level~\RNum{2}} \\
        Method & Par.~(M)~$\downarrow$ & GMACs~$\downarrow$ & PSNR~$\uparrow$ & SSIM~$\uparrow$ & PSNR~$\uparrow$ & SSIM~$\uparrow$ \\
        \midrule
        Degraded                                              &      &       & 36.68 & 0.939 & 35.39 & 0.937 \\
        
        UNet    
                                                              & 11.7 & \cellcolor{Red}19.5 & 37.46 & 0.958 & 36.35 & 0.953 \\
        DRUNet                                                & \cellcolor{Blue} 8.0  & 32.5  & 37.87 & 0.960 & 36.51 & 0.954 \\
        CycleISP                                              & \cellcolor{Red}2.8  & 192.3 & \cellcolor{Red}40.38 & \cellcolor{Red}0.968 & \cellcolor{Red}38.66 & \cellcolor{Red}0.965 \\
        Restormer~(*)                   & 26.1 & 132.4 & \cellcolor{Green}41.42 & \cellcolor{Blue}0.974 & \cellcolor{Green}39.75 & \cellcolor{Blue}0.972 \\
        NAFNet                          & 17.1 & \cellcolor{Blue}16.1 & 38.45 & 0.966 & 36.93 & 0.962 \\
        \midrule
        RawIR                                            & \cellcolor{Green}1.5  & \cellcolor{Green}\textbf{12.3} & \cellcolor{Blue}40.90 & \cellcolor{Green}0.974 & \cellcolor{Blue}38.89 & \cellcolor{Green}0.971 \\
        \bottomrule
    \end{tabular}
    }
    \caption{RAW Image Restoration Benchmark. We use the smartphone testset, and indicate the degradation level. We report metrics in the RAW domain. MACs calculated using a 248px input. (\textcolor{green}{Best}, \textcolor{cyan}{Second Best}, \textcolor{red}{Third Best}). Note that Restormer~\cite{zamir2022restormer} is notably more complex than the other models, and cannot handle high-resolution images.
    }
    \label{tab:benchmark}
\end{table}   

\begin{figure*}[t]
    \centering
    \includegraphics[width=\linewidth]{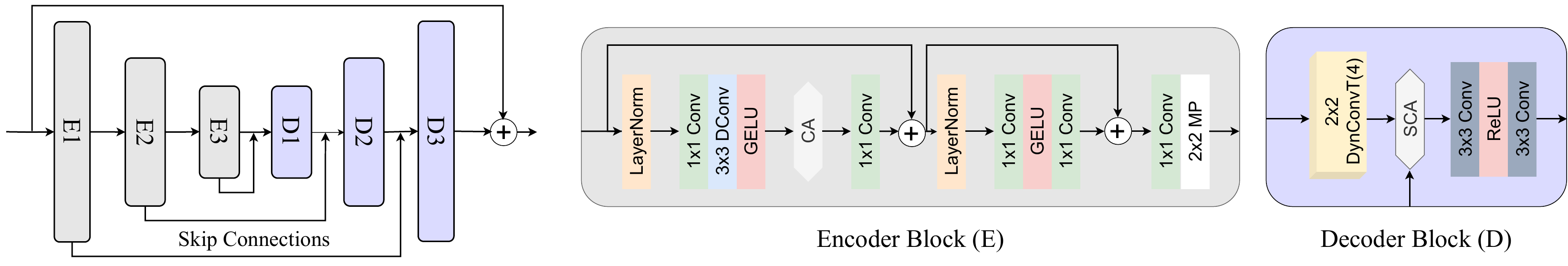}
    \caption{Illustration of RawIR, based on \emph{NAFNet}~\cite{chen2022simple}. We incorporate our designed Dynamic Convolution Block (\texttt{DynConvT})~\cite{verelst2020dynamic} into the decoder, and reduce the number of blocks. More details in the appendix.}
    \label{fig:rawir-arch}
\end{figure*}

\begin{figure}[t]
    \centering
    \setlength{\tabcolsep}{1.0pt}
    \begin{tabular}{cccc}
    \includegraphics[width=0.48\linewidth]{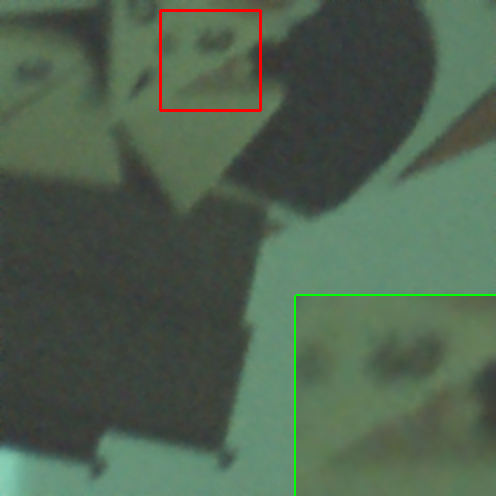} &
    \includegraphics[width=0.48\linewidth]{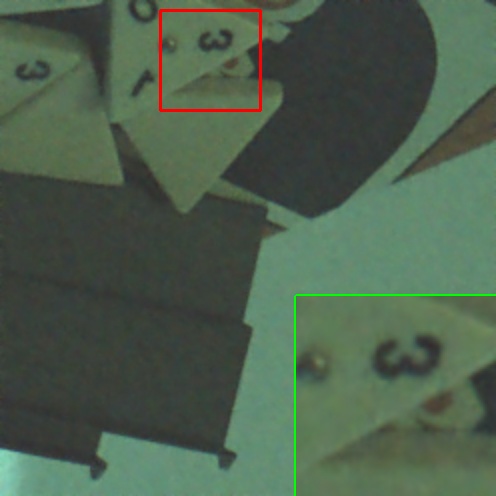}
    \tabularnewline
    Degraded & NAFNet \tabularnewline
    \includegraphics[width=0.48\linewidth]{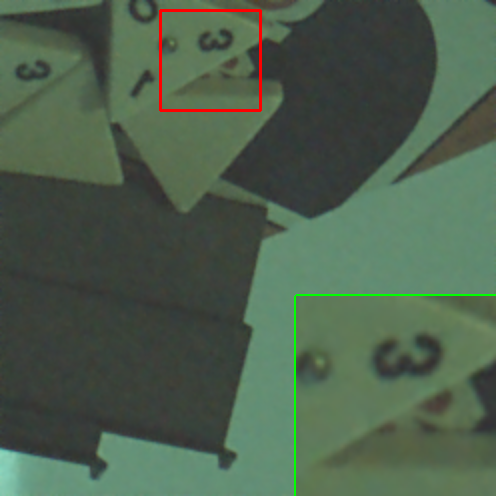} &
    \includegraphics[width=0.48\linewidth]{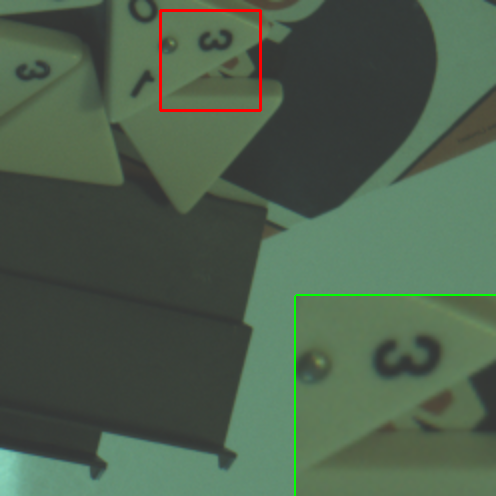} 
    \tabularnewline
    Ours RawIR & Ground-truth 
    \tabularnewline
    \end{tabular}
    \caption{Qualitative results of RAW restoration on a real-world capture (defocus and high ISO sample).
    }
    \label{fig:quali-results-real}
\end{figure}  

\subsection{RAW Restoration Benchmark} 

We build the benchmark around \emph{state-of-the-art} image restoration methods. Some of these methods are limited when processing HR images due to their complexity (\eg excessive MACs or parameters). For this reason, we propose \emph{RawIR} as an efficient baseline, able to process 4K resolution images in standard GPUs -120 ms for a 4K image on a RTX 3090Ti, without patching or tiling strategies. RawIR combines NAFNet~\cite{chen2022simple} baseline blocks with dynamic convolutions and simplified stereo channel attention in the decoder part of the model. The key components are: \textbf{i)} Inverted Residual Blocks combined with attention~\cite{howard2017mobilenets, chen2022simple}, learnt blending transposed convolutions -similar to Dynamic Convolutions for faster inference~\cite{verelst2020dynamic}- coupled with channel attention (SCA)~\cite{chen2022simple}. We illustrate the RawIR architecture in Figure~\ref{fig:rawir-arch}.

In comparison to NAFNet -already quite efficient- our method has $10\times$ less parameters and $\approx20\%$ less MACs. In comparison with Restormer~\cite{zamir2022restormer}, our method has $10\times$ less MACs. Our method is also notably faster than CycleISP~\cite{zamir2020cycleisp}.
We also consider UNet-like baselines, as this is a popular architecture in image restoraton. We study a classical deep residual UNet inspired in~\cite{brooks2019unprocessing}, and DRUNet~\cite{zhang2021DPIR}. Both represent a good approximation of the performance of other methods with similar architectures.

All the models were implemented in Pytorch, and trained using the same setup \ie Adam optimizer with learning rate $1e^{-4}$, $\mathcal{L}_1$ loss, and 200K iterations. The batch size depends on the model. We perform experiments using multiple NVIDIA RTX 3090Ti and 4090Ti.
Finally, we consider PSNR and SSIM as standard metrics in image restoration, including RAW image denosing~\cite{abdelhamed2018high}.

\vspace{3mm}

\begin{figure*}[!ht]
    \centering
    \setlength{\tabcolsep}{1.0pt}
    \begin{tabular}{cccccc}
    \includegraphics[width=0.2\linewidth]{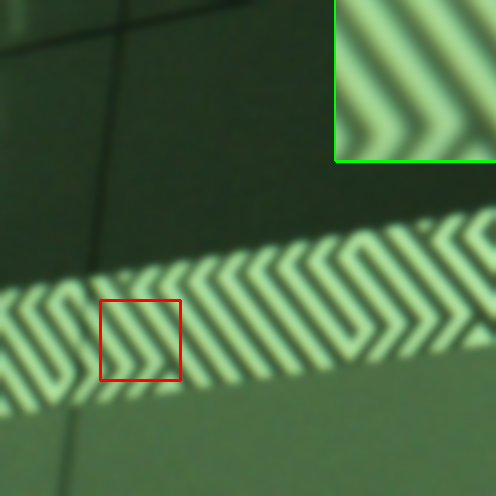} &
    \includegraphics[width=0.2\linewidth]{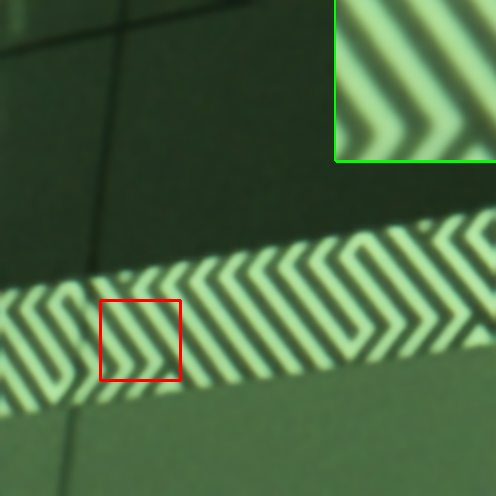} & 
    \includegraphics[width=0.2\linewidth]{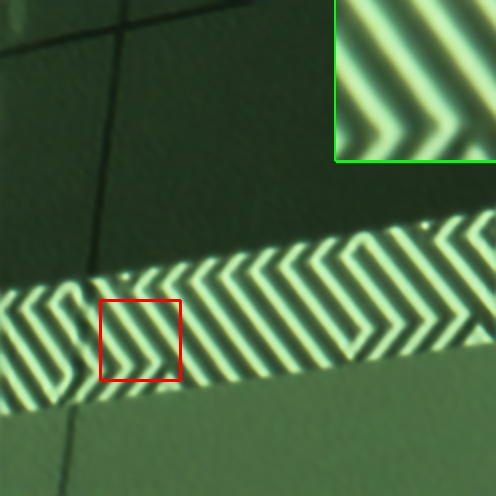} &
    \includegraphics[width=0.2\linewidth]{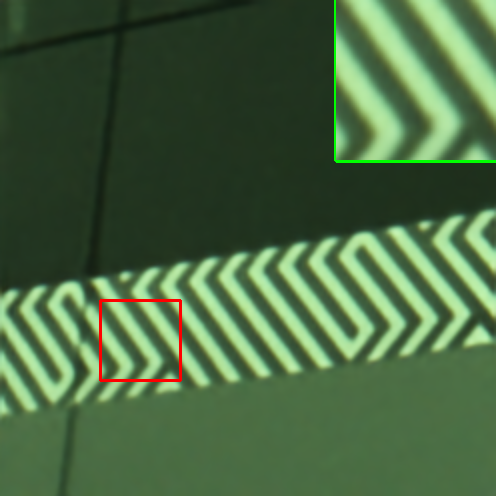} &
    \includegraphics[width=0.2\linewidth]{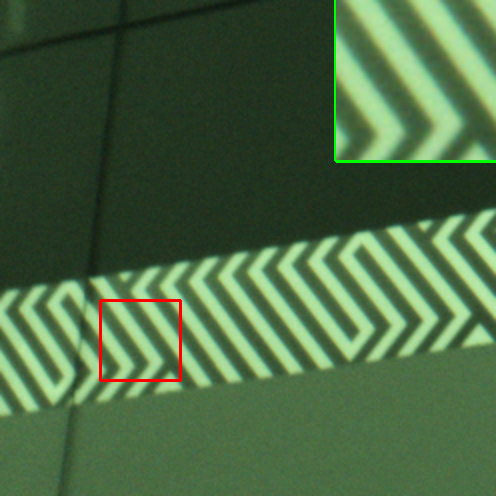}
    \tabularnewline
    \includegraphics[width=0.2\linewidth]{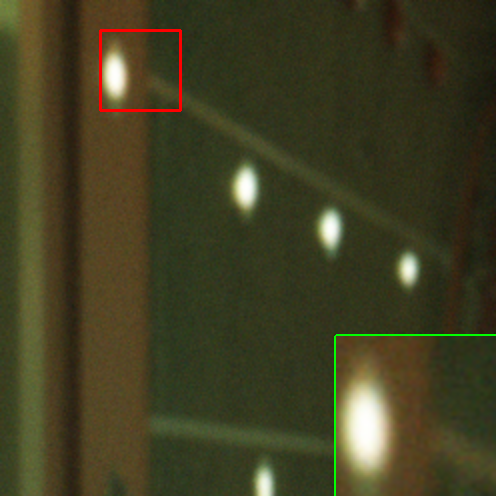} &
    \includegraphics[width=0.2\linewidth]{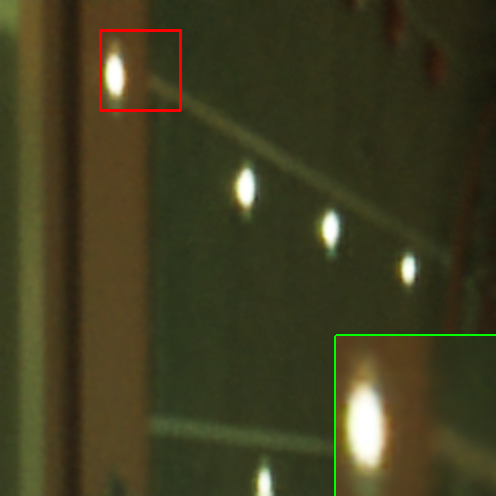} & 
    \includegraphics[width=0.2\linewidth]{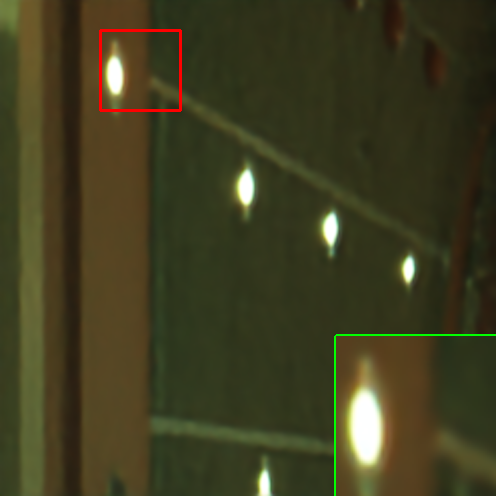} &
    \includegraphics[width=0.2\linewidth]{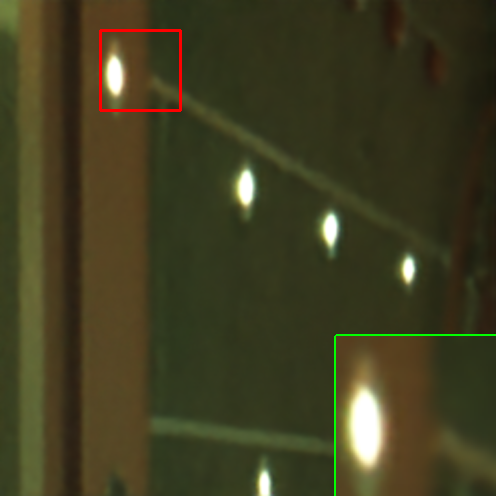} &
    \includegraphics[width=0.2\linewidth]{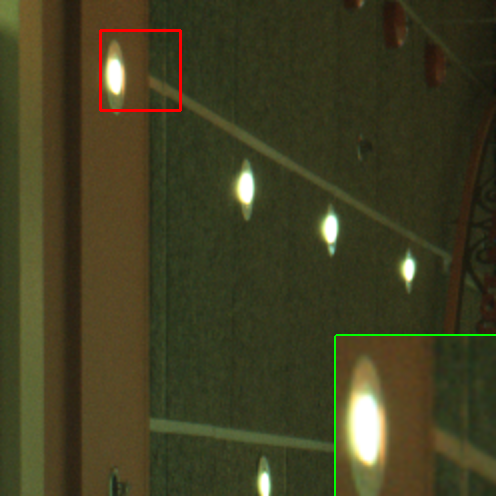}
    \tabularnewline
    \includegraphics[width=0.2\linewidth]{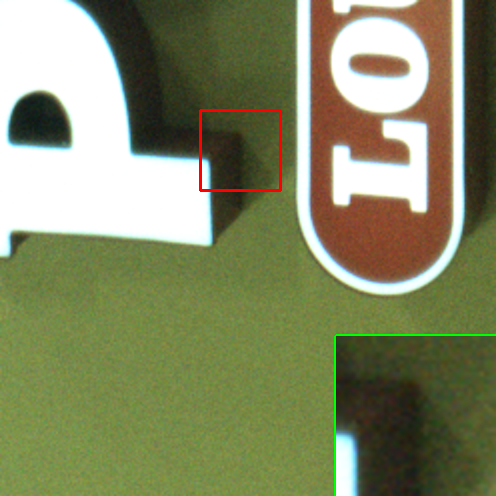} &
    \includegraphics[width=0.2\linewidth]{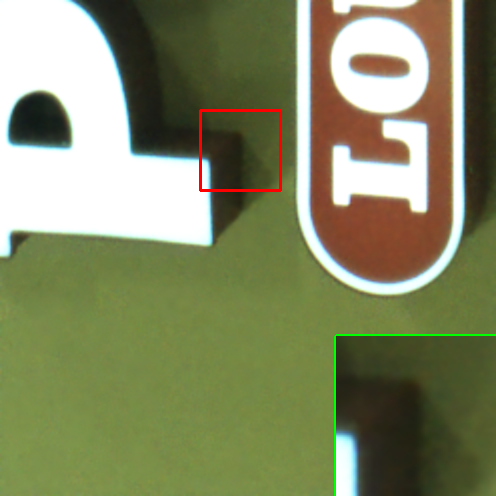} & 
    \includegraphics[width=0.2\linewidth]{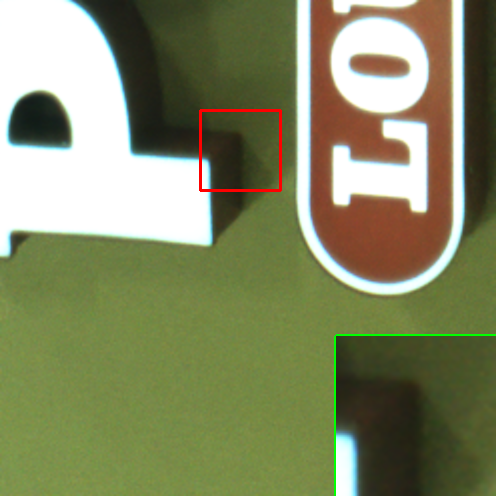} &
    \includegraphics[width=0.2\linewidth]{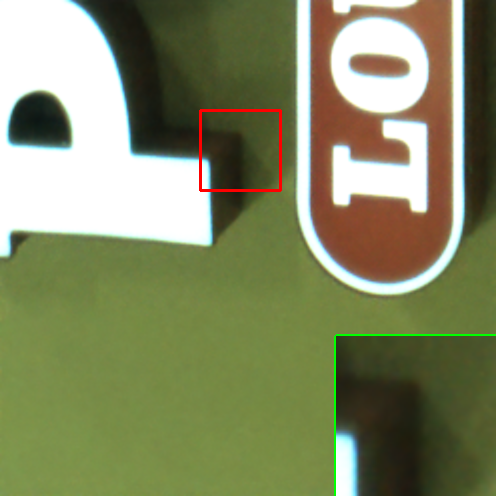} &
    \includegraphics[width=0.2\linewidth]{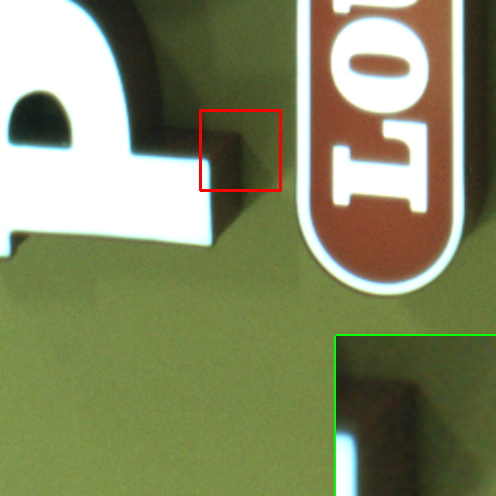}
    \tabularnewline
    \includegraphics[width=0.2\linewidth]{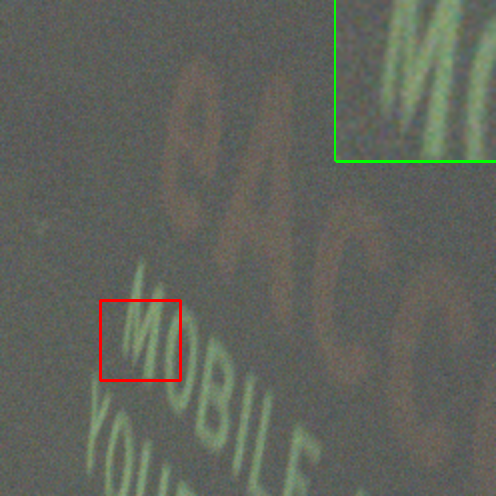} &
    \includegraphics[width=0.2\linewidth]{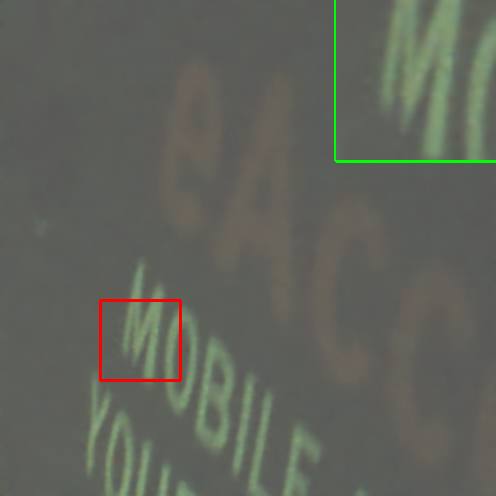} & 
    \includegraphics[width=0.2\linewidth]{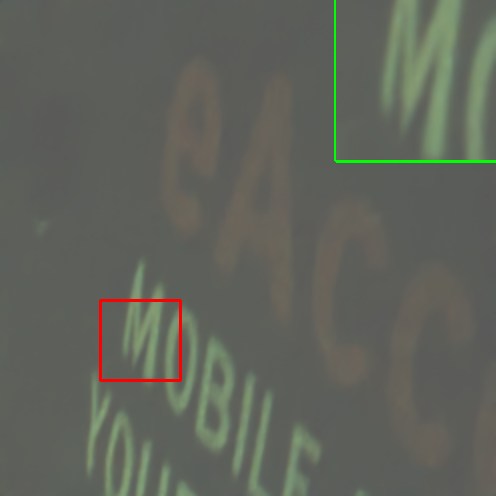} &
    \includegraphics[width=0.2\linewidth]{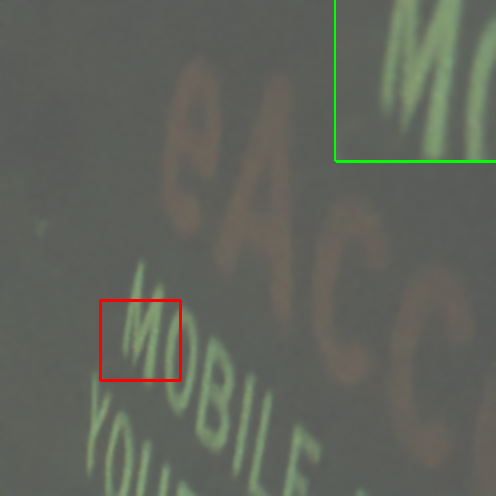} &
    \includegraphics[width=0.2\linewidth]{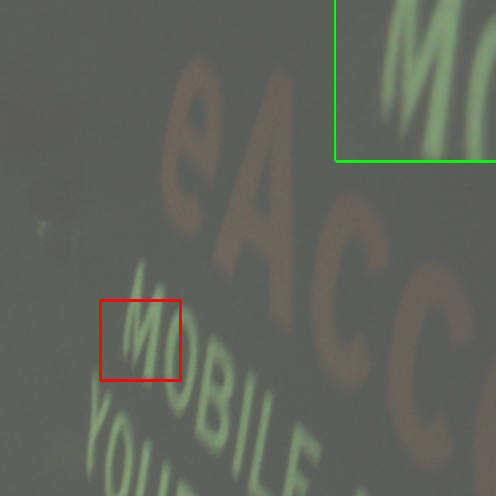}
    \tabularnewline
    Input & NAFNet~\cite{chen2022simple} & Restormer~\cite{zamir2022restormer} & RawIR (Ours) & Ground-truth
    \end{tabular}
    \caption{RAW restoration comparison using the smartphones testset. Our model can successfully restore RAW images with notable degradations, and it has $10\times$ less operations than Restormer~\cite{zamir2022restormer}. Best viewed in electronic version.}
    \label{fig:quali-results}
\end{figure*}

\noindent\textbf{Restoration results} In the \tref{tab:benchmark}, we provide a quantitative evaluation of multiple models for RAW image restoration. The models were trained using 22000 smartphone RAW images that are degraded on-line using our degradation pipeline \ie level \RNum{2}.
The synthetic test dataset consists of 2100 smartphone RAW images (see Table~\ref{tab:datasets}) degraded using our degradation pipeline levels \RNum{1} and \RNum{2} independently, and thus leading to two different testsets. Our model RawIR achieves great results while being extremely fast. 
We find that NAFNet and Restormer -both studied on RGB images- might require to be further adapted for RAW data, as the number of channels increases (from 3 to 4), and the signal properties are different. We provide \emph{qualitative results} in Figures~\ref{fig:quali-results} and~\ref{fig:quali-results-real}. Models can correct complex degradations and produce reasonable restored RAWs.

Figure~\ref{fig:quali-results} qualitative results on a real-world capture show the models are able to produce noise-free sharp images with proper textures and color distributions.


\noindent\textbf{Limitations~} Despite the overall promising results, we acknowledge certain limitations. The model struggles specially in low-light or underexposed scenes. Better modelling using device-specific PSFs (estimated via ray-tracing), and low-light noise profiles could make the model more robust.

\section{Conclusion}
\label{sec:conclusion}

In this paper we discuss a controllable degradation pipeline to synthesize realistic degraded RAW images for training deep blind restoration models.
We have curated a dataset with different smartphone sensors, and we also set a new benchmark for blind RAW restoration. 
Our experiments demonstrate that models trained with our degradation pipeline can restore directly sensor RAW images. This represents a powerful alternative for blind image restoration in the wild, and can be of benefit to other low-level downstream tasks. The code and dataset will be released upon acceptance.

\vspace{2mm}
\noindent\textbf{Other Works} We also investigate super-resolution in the RAW domain (BSRAW~\cite{conde2024bsraw}) to complement this work.

\vspace{2mm}
\noindent\textbf{Acknowledgments} This work was partially supported by the Humboldt Foundation (AvH).


\small{
\bibliographystyle{IEEEbib}
\bibliography{egbib}
}

\end{document}